\documentclass[11pt]{article}
\usepackage{times}
\usepackage{pdfpages}
\usepackage{setspace}
\usepackage{enumitem}
\usepackage{float}

\usepackage{wrapfig}
\usepackage{lipsum}
\usepackage{graphicx}

\usepackage{array}
\usepackage{subfigure}
\usepackage{booktabs}
\usepackage{multicol}
\usepackage{multirow}
\usepackage{threeparttable}
\usepackage[small, bf]{caption}
\usepackage[tableposition=top]{caption}
\usepackage{booktabs}
\usepackage{siunitx}

\usepackage{pgfgantt}

\usepackage{wrapfig}
\usepackage{times}
\usepackage{url}
\usepackage{soul}
\usepackage{color}
\begingroup
\makeatletter
\g@addto@macro{\UrlSpecials}{%
  \endlinechar=13 \catcode\endlinechar=12
  \do\%{\Url@percent}\do\^^M{\break}}
 \gdef\Url@percent{\@ifnextchar^^M{\@gobble}{\mathbin{\mathchar`\%}}}%
\endgroup %
\usepackage{array}
\usepackage{subfigure}

\usepackage{array}
\newcolumntype{L}[1]{>{\raggedright\let\newline\\\arraybackslash\hspace{0pt}}m{#1}}

\def\firstitem#1{\addtolength{\itemsep}{-0.5\baselineskip}
\item{\vskip -.10in #1}}

\usepackage[headheight=14pt,tmargin=1in,bmargin=1in,lmargin=1in,rmargin=1in]{geometry}

\usepackage{fancyhdr}
\usepackage{datetime}
\usepackage{indentfirst}
\usepackage[sort&compress, square, comma, numbers]{natbib}

\usepackage[compact]{titlesec}

\usepackage{titlesec}
\titlespacing*{\section}{0pt}{0.1\baselineskip}{0.1\baselineskip}
\titlespacing*{\subsection}{0pt}{0.2\baselineskip}{0.1\baselineskip}
\titlespacing*{\subsubsection}{0pt}{0.2\baselineskip}{0.1\baselineskip}
\titlespacing*{\paragraph}{0pt}{0.2\baselineskip}{0.2\baselineskip}

\usepackage{blindtext}

\usepackage{threeparttable}
\usepackage{soul}
\usepackage{booktabs}
\usepackage{multirow}
\usepackage{listings}

\definecolor{backcolour}{RGB}{246, 246, 246}   
\definecolor{codegreen}{RGB}{16, 124, 2}       
\definecolor{codepurple}{RGB}{170, 0, 217}     
\definecolor{codered}{RGB}{154, 0, 18}         

\lstdefinestyle{gcolabstyle}{
  basicstyle=\ttfamily\small,
  backgroundcolor=\color{backcolour},   
  commentstyle=\itshape\color{codegreen},
  keywordstyle=\color{codepurple},
  stringstyle=\color{codered},
  numberstyle=\ttfamily\footnotesize\color{darkgray}, 
  breakatwhitespace=false,         
  breaklines=true,                 
  captionpos=b,                    
  keepspaces=true,                 
  numbers=left,                    
  numbersep=5pt,                  
  showspaces=false,                
  showstringspaces=false,
  showtabs=false,                  
  tabsize=2
}

\begin{document}

\thispagestyle{empty}
\begin{center}
{\bf \Large Examining the Challenges in Archiving Instagram}\\

\vspace{5mm}
Rachel Zheng, rzheng02@wm.edu\\
Department of Computer Science, College of William and Mary\\
\hfill \break
Michele C. Weigle, mweigle@cs.odu.edu\\
Department of Computer Science, Old Dominion University\\
\end{center}

\begin{abstract}
To prevent the spread of disinformation on Instagram, we need to study the accounts and content of disinformation actors. However, due to their malicious nature, Instagram often bans accounts that are responsible for spreading disinformation, making these accounts inaccessible from the live web. The only way we can study the content of banned accounts is through public web archives such as the Internet Archive. However, there are many issues present with archiving Instagram pages. Specifically, we focused on the issue that many Wayback Machine Instagram mementos redirect to the Instagram login page. In this study, we determined that mementos of Instagram account pages on the Wayback Machine began redirecting to the Instagram login page in August 2019. We also found that Instagram mementos on Archive.today, Arquivo.pt, and Perma.cc are also not well archived in terms of quantity and quality. Moreover, we were unsuccessful in all our attempts to archive Katy Perry's Instagram account page on Archive.today, Arquivo.pt, and Conifer. Although in the minority, replayable Instagram mementos exist in public archives and contain valuable data for studying disinformation on Instagram. With that in mind, we developed a Python script to web scrape Instagram mementos. As of August 2023, the Python script can scrape Wayback Machine archives of Instagram account pages between November 7, 2012 and June 8, 2018. 
\end{abstract}

\section{Introduction}
As of January 2023, Instagram was the fourth most popular social media app worldwide, boasting over 2 billion active monthly users \cite{instagram_popularity}. Instagram's large user base makes it an attractive tool for malicious actors to spread disinformation. For example, the Russian Internet Research Agency (IRA) used Instagram, Twitter, and Facebook to manipulate narratives on Americans culture and politics leading up to the 2016 US presidential election. Out of the three social media platforms, Instagram had the most disinformation engagement \cite{US_senate_report}. However, while just as prevalent, disinformation on Instagram is understudied as compared to disinformation on Facebook and Twitter. Our overall goal for this study is to contribute to disinformation research on Instagram, specifically by examining archived Instagram pages. Archived versions, or mementos, of Instagram account pages allow us to view what an Instagram page looked like in the past and track changes to a Instagram page. Instagram mementos are especially important in disinformation research as accounts and posts that spread disinformation are often made unavailable on live Instagram. Public web archives are the only place we can still see any deleted or banned disinformation content. In a project from the 2022 NSF REU Site in Disinformation Detection and Analytics, Haley Bragg \cite{bragg-reu-arxiv, bragg-jcdl23} examined the mementos of the Instagram account pages of the Disinformation Dozen, 12 individuals identified by the Center of Countering Digital Hate to be responsible for 65\% of anti-vaccine information on social media \cite{CCDH_disinfo_dozen}. Bragg found that upon replay, most of the Instagram mementos redirected to the Instagram login page, making it difficult to examine the Instagram pages. 

\begin{figure*}[ht]
\centering
\includegraphics[width=.9\linewidth]{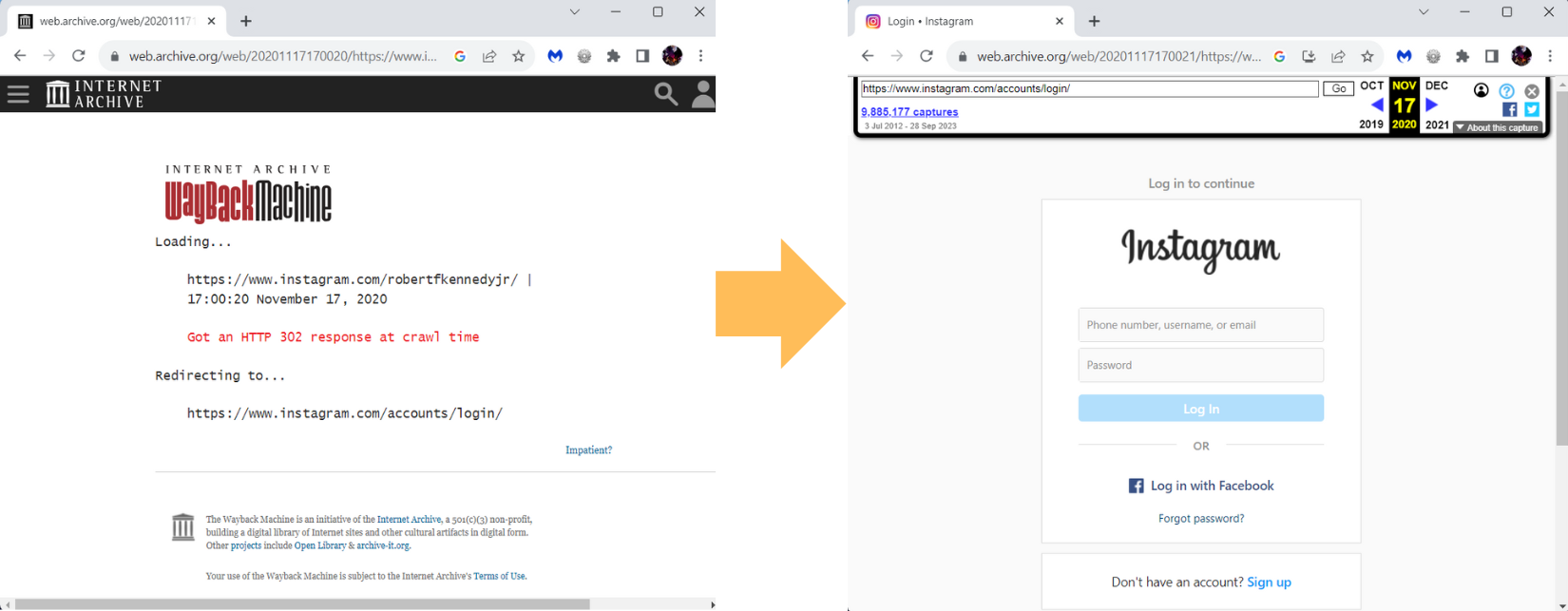}
\caption{Redirection to Instagram login page}
\label{fig:redirect_to_login}
\end{figure*} 

Motivated by Bragg's work, we set out to pinpoint when redirects to the login page became prominent. We also examined the mementos of Instagram pages on Archive.today, Arquivo.pt, and Perma.cc. We attempted to archive Instagram pages using the Wayback Machine, Archive.today, Arquivo.pt, and Conifer. Finally, we developed a Python script to scrape information from Instagram mementos on the Wayback Machine. The code and analysis is available in a GitHub repository.\footnote{\url{https://github.com/oduwsdl/Zheng-REU-2023}}

\section{Related Work}
In November 2020, Himarsha Jayanetti conducted the study “How well is Instagram archived?” \cite{jayanetti_2020} and found that Instagram was not well archived. In fact, compared to other popular social media platforms such as Facebook and Twitter, Instagram was significantly underperforming in both its archive quantity and quality. In her study, Jayanetti focused on Katy Perry’s Instagram account, as Perry was one of the most followed Instagram account in October 2020, with over 100 million followers. Furthermore, she noticed that Katy Perry posted the same content across her Instagram, Facebook, and Twitter accounts. This detail ensured that any possible bias from the type or quantity of posts was minimized when comparing how well Katy Perry’s accounts were archived across all three social media platforms. Using MemGator \cite{jcdl-2016:alam:memgator}, a tool that retrieves the TimeMap for a given URL from 16 different archives, Jayanetti observed that Katy Perry’s Instagram page was the least archived (Table \ref{table:katy_perry_archives_across_social_media}), indicating that Instagram is not as well archived as Facebook or Twitter. 
\begin{table}[ht]
    \center
    \begin{tabular}{|c|c|c|c|}
    \hline
    &Instagram&Facebook&Twitter\\
    \hline
    Internet Archive & 1803 & 2032 & 4025\\
    Archive-It & 108 & 1234 & 1185\\
    archive.today & 7 & 0 & 2\\
    Library of Congress & 4 & 1 & 0\\
    UK Web Archive & 0 & 16 & 58\\
    Australian Web Archive & 0 & 37 & 2\\
    Portuguese Web Archive & 2 & 0 & 1\\
    \hline 
    Total & 1924 & 3320 & 5273\\
    \hline
    \end{tabular}
    \caption{Captures of Katy Perry’s social media account pages across different public web archives. Table reproduced from \cite{jayanetti_2020}.}\label{table:katy_perry_archives_across_social_media}
\end{table}

In addition, only 31.67\% of Katy Perry’s Instagram posts were archived in public web archives. Jayanetti also noted that the Instagram post URLs are completely opaque, meaning that one could not tell who the posts belong to by just looking at the URL. This is not the case for Facebook and Twitter URLs, which contain the username in the URL. With Facebook and Twitter posts, one could do a prefix search using Internet Archive’s Wayback CDX Server API to search for all the posts that belong to a specific account. However, due to the completely opaque nature of Instagram post URLs, a prefix search is currently not feasible. This makes searching for all the posts of a specific account difficult \cite{jayanetti_thesis}. Lastly, Jayanetti observed many quality and replayability issues with Instagram mementos. If popular Instagram accounts such as Katy Perry’s were not archived well, it leaves the question of how much worse less popular Instagram accounts were archived.

\begin{figure*}[ht]
\centering
\includegraphics[width=.9\linewidth]{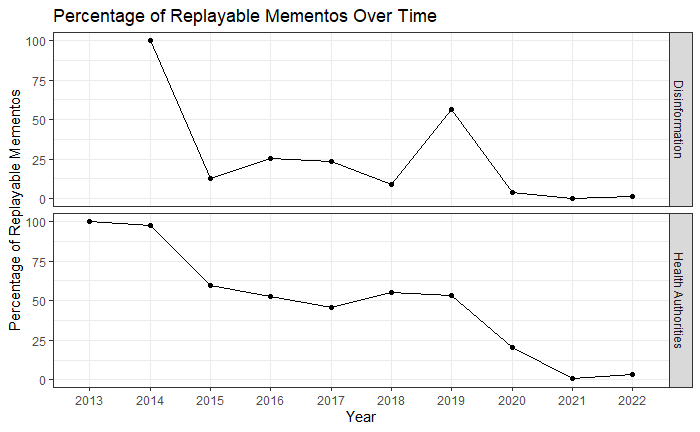}
\caption{Percentage of Replayable Mementos Over Time. Reproduced from \protect\cite{bragg-reu-arxiv}.}
\label{fig:haley_percentage_replayable_graph}
\end{figure*}

Looking more in depth at the replayability of Instagram mementos, Bragg \cite{bragg-reu-arxiv, bragg-jcdl23} found that many Instagram account page mementos redirected to the Instagram login page. Even if a memento did not redirect to the login page, there was no guarantee that all post images would replay. In Bragg’s study, she collected the Instagram account page mementos of two main groups, the Disinformation Dozen and a set of known health authorities. First, Bragg determined what percentage of the mementos were replayable and observed that there were two significant drops in the percentage of replayable mementos, one from 2014-2015 and the other from 2019-2020, as shown in Figure \ref{fig:haley_percentage_replayable_graph}. By 2022, only 1.23\% of mementos for the Disinformation Dozen and 3.34\% of the mementos for the health authorities were replayable. Out of the replayable mementos, over half were missing at least one post image. In fact, for the Disinformation Dozen, a majority of their Instagram mementos had no post images at all. This is especially concerning as most of the Disinformation Dozen have been banned from Instagram, making public archives the only place that their posts can still be viewed for researching their methods of spreading disinformation.

\section{Pinpointing When Redirects to the Login Page Became Prominent}
As a continuation to Bragg’s work, we wanted to pinpoint when redirects to the Instagram login page became prominent. To do this, we repeated Bragg’s memento replayability analysis with the top 25 most followed Instagram accounts (Table \ref{table:top_25_most_followed_instagram_accounts}), gathered from Wikipedia.\footnote{\url{https://en.wikipedia.org/w/index.php?title=List_of_most-followed_Instagram_accounts&oldid=1169463870}} We chose to use this specific set of Instagram accounts to get a larger dataset of mementos that was more representative of a typical Instagram user. Using Internet Archive’s Wayback CDX Server API,\footnote{\url{https://github.com/internetarchive/wayback/tree/master/wayback-cdx-server}} we collected data for 75,378 Instagram account page mementos from the Wayback Machine. By default, the CDX API gives the timestamp, original URI (URI-R), MIME type, and status code for each memento of a given URL. We classified mementos with a 200 status code as successes, a 3xx status code as redirects, and a 4xx or 5xx status code as errors. Some mementos also had a status code of ``-” and a MIME type of warc/revisit. In most cases, a memento is a warc/revisit if the webpage at the time of archiving was a complete or near duplicate of another memento. In addition, we also retrieved the final URI for any redirects and warc/revisit mementos. 
\begin{table}[ht!]
    \center
    \begin{tabular}{|c|c|r|}
    \hline
    Name & Instagram Handle & Mementos\\
    \hline
    Instagram & instagram & 4981\\
    Christiano Ronaldo & cristiano & 1913\\
    Lionel Messi & leomessi & 2294\\
    Selena Gomez & selenagomez & 3712\\
    Kylie Jenner & kyliejenner & 5113\\
    Dwayne Johnson & therock & 2354\\
    Ariana Grande & arianagrande & 2402\\
    Kim Kardashian & kimkardashian & 8883\\
    Beyonce & beyonce & 3582\\
    Khloe Kardashian & khloekardashian & 3081\\
    Nike & nike & 1169\\
    Kendall Jenner & kendalljenner & 3419\\
    Justin Bieber & justinbieber & 4338\\
    National Geographic & natgeo & 4719\\
    Taylor Swift & taylorswift & 3351\\
    Virat Kohli & virat.kohli & 291\\
    Jennifer Lopez & jlo & 3541\\
    Nicki Minaj & nickiminaj & 2460\\
    Kourtney Kardashian & kourtneykardash & 3564\\
    Miley Cyrus & mileycyrus & 3581\\
    Neymar & neymarjr & 1073\\
    Katy Perry & katyperry & 2625\\
    Zendaya & zendaya & 920\\
    Kevin Hart & kevinhart4real & 813\\
    Cardi B & iamcardib & 1199\\
    \hline 
    \end{tabular}
    \caption{Instagram handles and the number of mementos (as of June 2023) of the top 25 most followed Instagram accounts}
    \label{table:top_25_most_followed_instagram_accounts}
\end{table}

\begin{figure*}
\centering
\includegraphics[width=.9\linewidth]{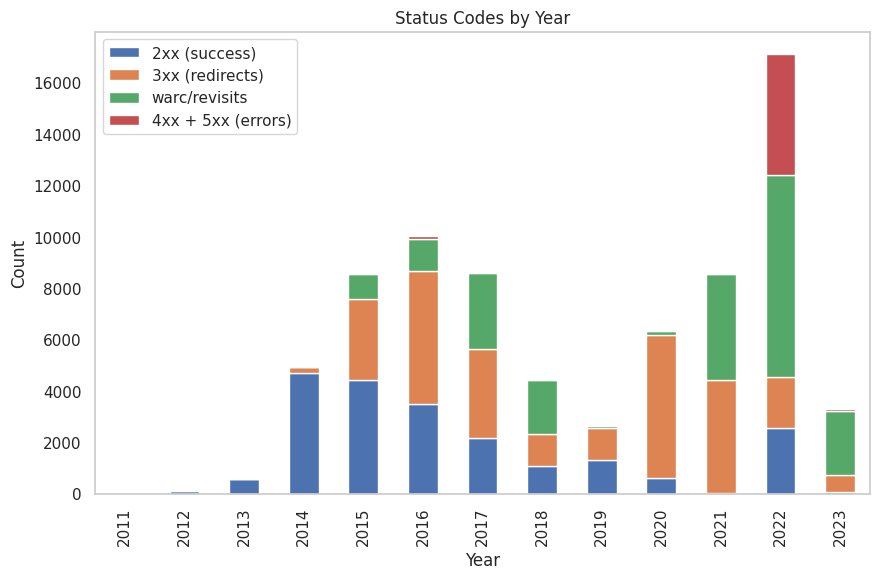}
\caption{Distribution of status codes of the mementos of the the top 25 most followed Instagram accounts over time}
\label{fig:top_25_status_code_distribution}
\end{figure*}

From an initial look at the data, we noticed that many redirects prior to 2019 did not redirect to the Instagram login page, but instead redirected to another URI-M. These redirects occur due to URI canonicalization of the URI-R and are not an accurate representation of the URI-R at its given timestamp \cite{kelly-arxiv17}. The dramatic increase in redirects from 2014 to 2015 (see Figure \ref{fig:top_25_status_code_distribution}) can be explained by a scheme change from http to https that began in February 2015. This is likely the reason for the sharp decline from 2014-2015 in the percentage of replayable mementos as seen in Figure \ref{fig:haley_percentage_replayable_graph}. Since we only wanted to consider unique representations of the Instagram mementos for this study, we decided to use a different formula from Bragg in calculating the percentage of mementos that were replayable:
\begin{equation}
    \label{eq:new eq for percentage replayable}
    \text{percentage replayable} = \frac{\text{200s + warc/revisits to 200s}}
                                {\text{200s + warc/revisits + 3xx to login + 4xx + 5xx}} * 100
\end{equation}
Our formula for percentage of replayable mementos differs in the denominator we used. Instead of using the total number of mementos returned by the CDX API, we chose to count only the unique representations of the mementos in the denominator. In other words, we did not count any redirects, except for those that redirected to the Instagram login page.

\begin{figure*}[ht]
\centering
\includegraphics[width=.9\linewidth]{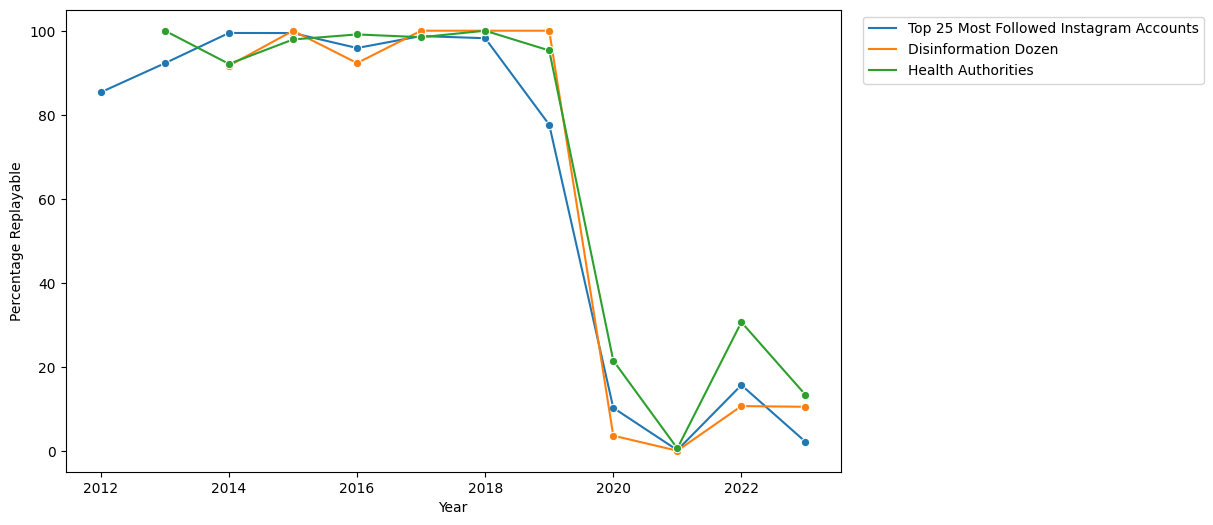}
\caption{Percentage of Replayable Mementos Over Time}
\label{fig:percentage_replayable_3_datasets}
\end{figure*}

\begin{figure*}[ht]
\centering
\includegraphics[width=.45\linewidth]{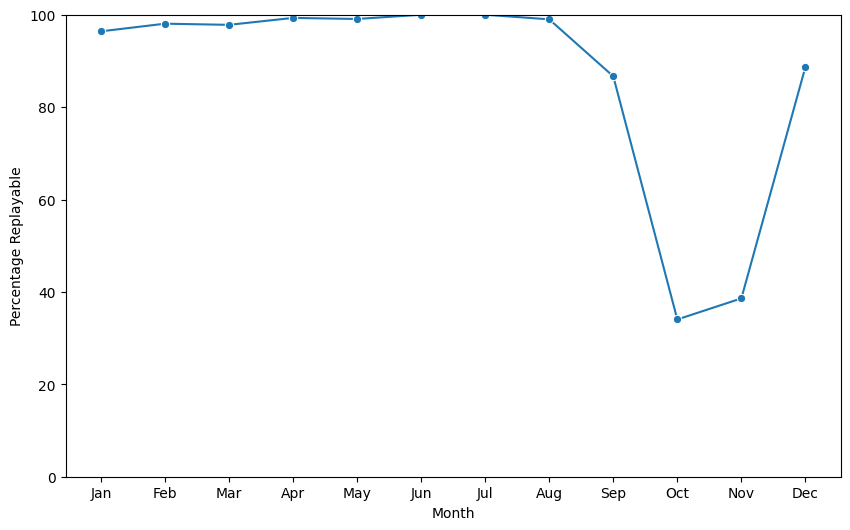}
\includegraphics[width=.45\linewidth]{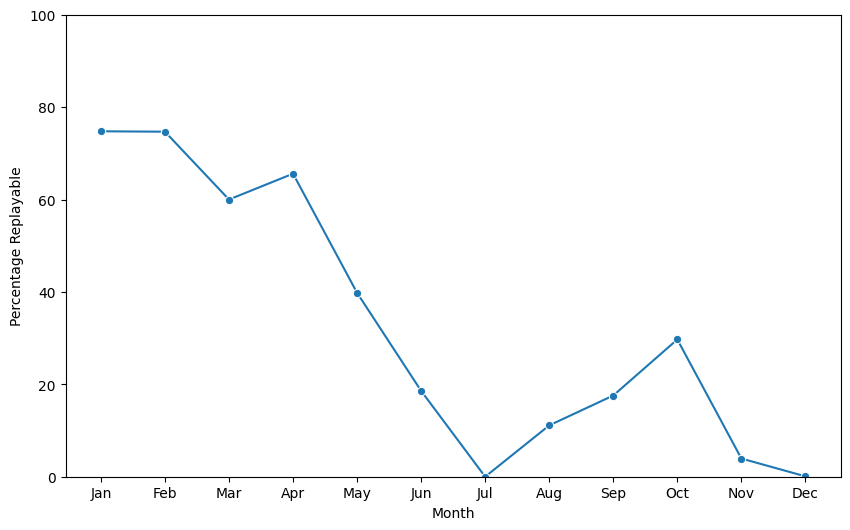}
\caption{Percentage of Replayable Mementos in 2019 (left) and 2020 (right)}
\label{fig:percentage_replayable_2019_2020}
\end{figure*}

Figure \ref{fig:percentage_replayable_3_datasets} shows that all three data sets of Instagram accounts have similar patterns. This indicates that any trend in percentage replayable was the result of an Instagram-wide occurrence and not specific to the accounts of any one field. Most noticeably, the graph shows a sharp decline from 2018-2020 in the percentage of replayable mementos. Taking a closer look at 2019 and 2020, we can see in Figure \ref{fig:percentage_replayable_2019_2020} that the decline in percentage replayable really started around August/September 2019, and by late 2020 the percentage of replayable mementos was consistently below 40\%.
These observations led us to suspect that redirects to the login page became prominent around August 2019.
 However, increases in errors, warc/revisits to errors, and/or revisits to errors could also be the cause of the decline in percentage of replayable mementos. 

\begin{figure*}[ht]
\centering
\includegraphics[width=.75\linewidth]{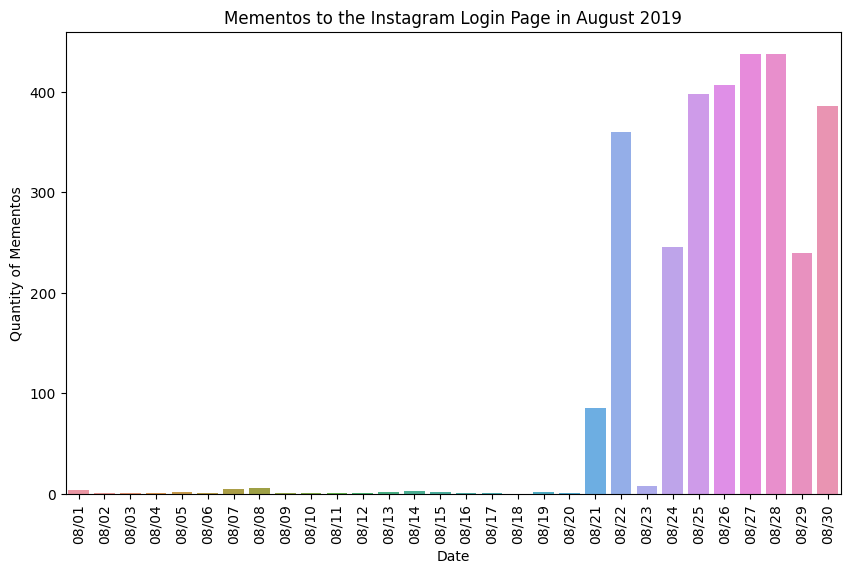}
\caption{Mementos to the Instagram login page in August 2019}
\label{fig:mementos_to_instagram_login_2019}
\end{figure*}

To further support that redirection to the login page was the primary reason for the massive drop in percentage replayable, we also retrieved the CDX data for mementos of the Instagram login page itself, \url{www.instagram.com/accounts/login}. We wanted to see if there were any trends over time in the amount of mementos to the Instagram login page that might indicate the beginning of redirections to the login page.
Figure \ref{fig:mementos_to_instagram_login_2019} shows a significant increase in the count of mementos to the login page from August 20, 2019 to August 21, 2019. Moreover, the first redirect to the login page from all three data sets (Disinformation Dozen, Health Authorities, and Top 25 Most Followed Instagram Accounts) was a memento of Instagram account \emph{thisisbillgates} on August 22, 2019. All of this has led us to conclude that redirections to the Instagram login page from Instagram mementos began in August 2019. This change in Instagram mementos occurred around the same time Instagram started limiting the amount of posts users could see without logging in \cite{instagram_login_policy_1, instagram_login_policy_2}. 

\section{Looking at Instagram Mementos from Other Archives}
So far, our research, as well as the previously mentioned studies, mainly focused on mementos from the Wayback Machine. However, we were curious to see if other archiving services had similar replayability and quality issues. We looked at existing mementos from Archive.today, Arquivo.pt, and Perma.cc, specifically focusing on the mementos around 2019, the year redirects to the Instagram login page started occurring on the Wayback Machine. There were many issues present in Instagram mementos on Archive.today, Arquivo.pt, and Perma.cc.

\subsection{Archive.today}
\begin{figure*}[ht]
\centering
\includegraphics[width=.46\linewidth]{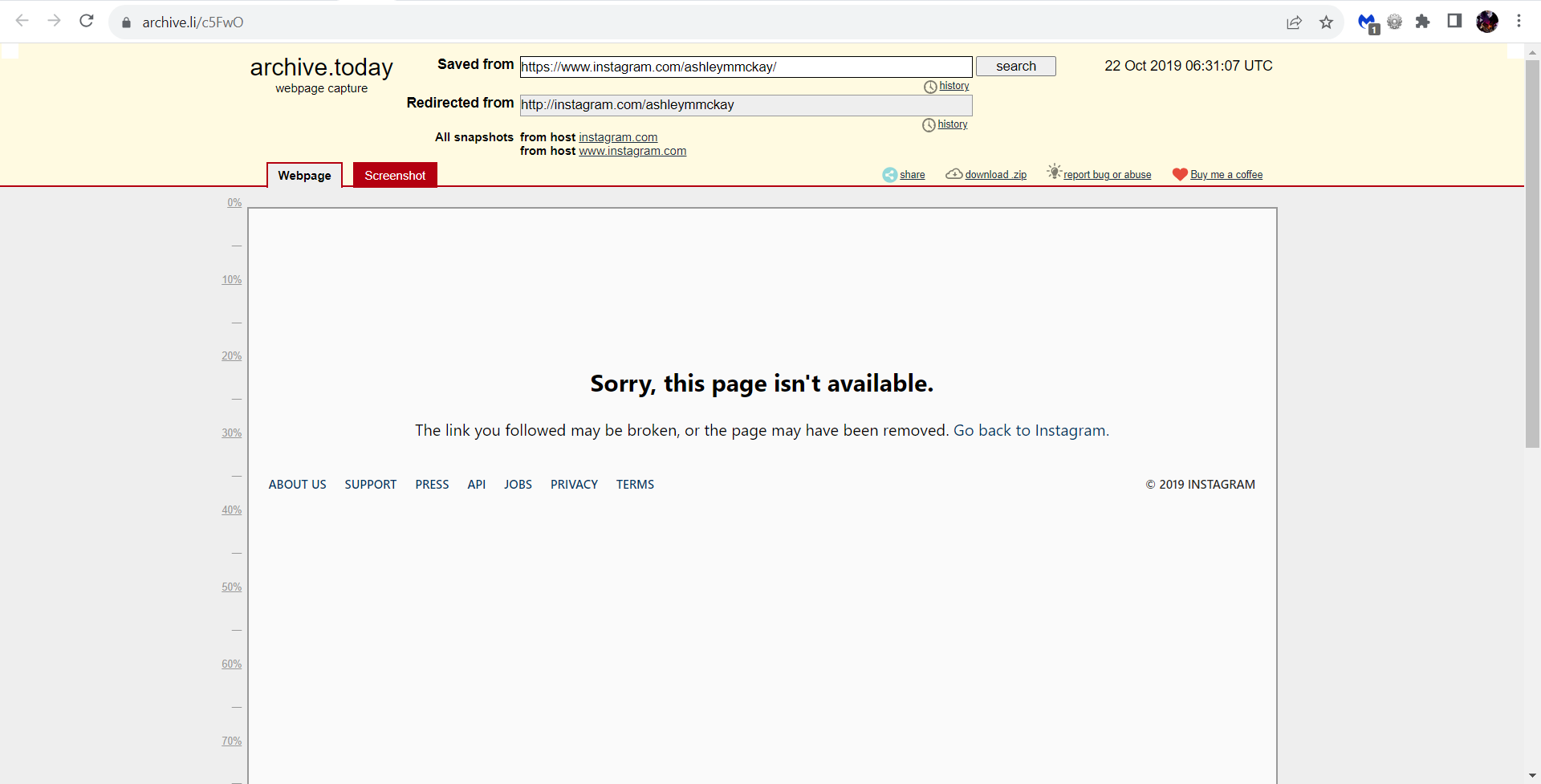}
\includegraphics[width=.46\linewidth]{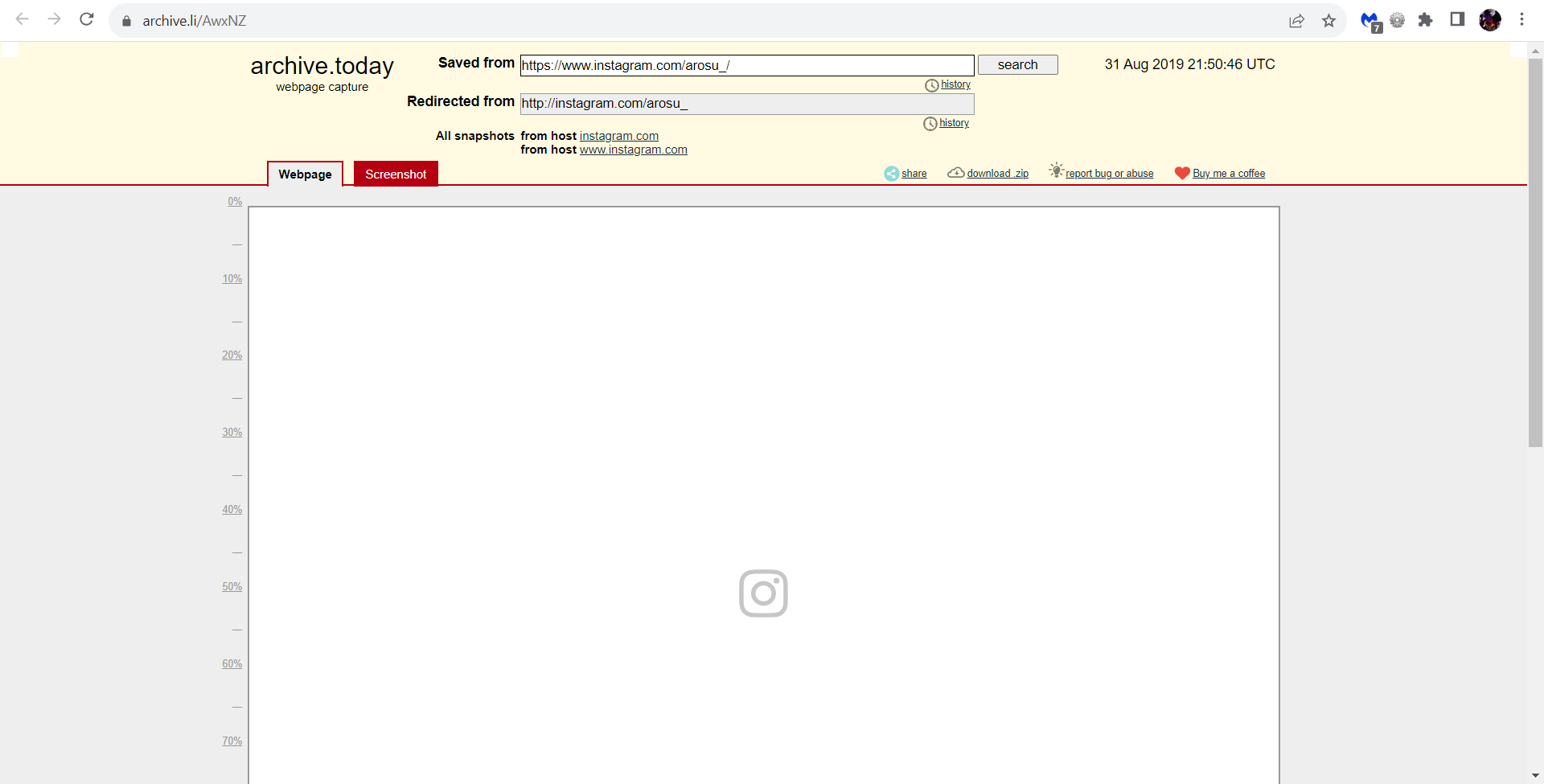}
\caption{Examples of Archive.today Instagram Mementos from 2018-Nov 2019}
\label{fig:archive_today_mementos}
\end{figure*}

Compared to the Wayback Machine, there are significantly fewer mementos of the top 25 most followed Instagram account on Archive.today and even fewer from 2019. Since Archive.today does not support the CDX Server API, we chose to use Archive.today's prefix search feature to view more Instagram mementos from 2019. We did a prefix search with http://instagram.com/a*, http://instagram.com/b*, http://instagram.com/c*, etc. to get a set of Archive.today Instagram mementos to analyze. From the 1,698 mementos from 2018 to November 2019 that we viewed, we were unable to see any information relevant to the specific Instagram account in these mementos. Instead, we found mementos with 404 errors, completely blank mementos, or mementos with only the Instagram logo (Figure \ref{fig:archive_today_mementos}). 

\subsection{Arquivo.pt}
\begin{figure*}[ht]
\centering
\includegraphics[width=.9\linewidth]{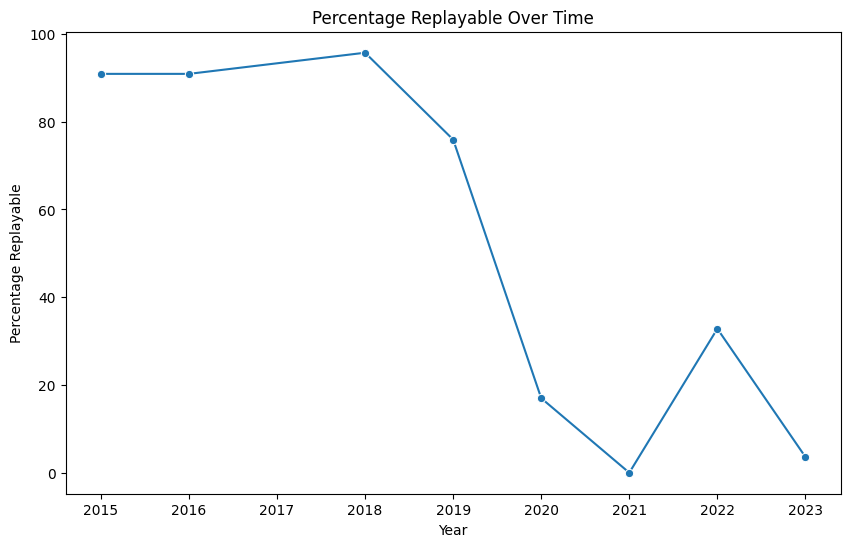}
\caption{Percentage of Replayable Mementos Over Time for the Mementos of the Top 25 most followed Instagram accounts in Arquivo.pt}
\label{fig:percentage_replayable_arquivo}
\end{figure*}

\begin{figure*}[ht]
\centering
\includegraphics[width=.75\linewidth]{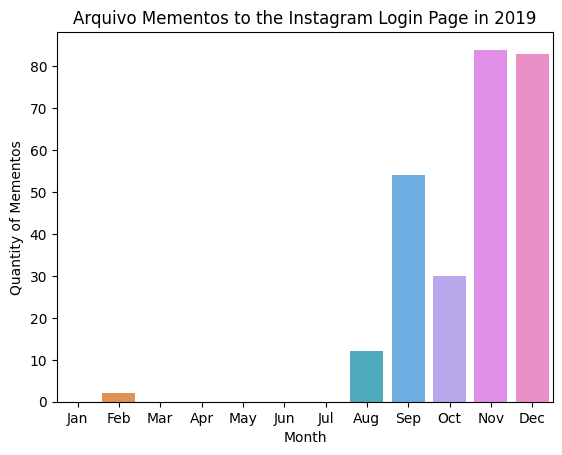}
\caption{Arquivo.pt mementos to the Instagram login page in 2019}
\label{fig:mementos_to_instagram_login_2019_arquivo}
\end{figure*}

Since Arquivo.pt supports the CDX Server API, we decided to do the same memento replayability analysis (see Section 3.1) on the 1,682 mementos of the top 25 most followed Instagram accounts from Arquivo.pt. In addition, we also retrieved mementos to www.instagram.com/accounts/login. Similar to the Wayback Machine, Instagram mementos in 2019 were heavily impacted by the start of redirections to the Instagram login page. This can be seen from similar patterns in the percentage replayable over time of the mementos of the top 25 most followed Instagram accounts (Figure \ref{fig:percentage_replayable_arquivo}) as well as the rise in the quantity of mementos to the Instagram login page in August 2019 (Figure \ref{fig:mementos_to_instagram_login_2019_arquivo}).

\subsection{Perma.cc}
\begin{figure*}[ht]
\centering
\includegraphics[width=.75\linewidth]{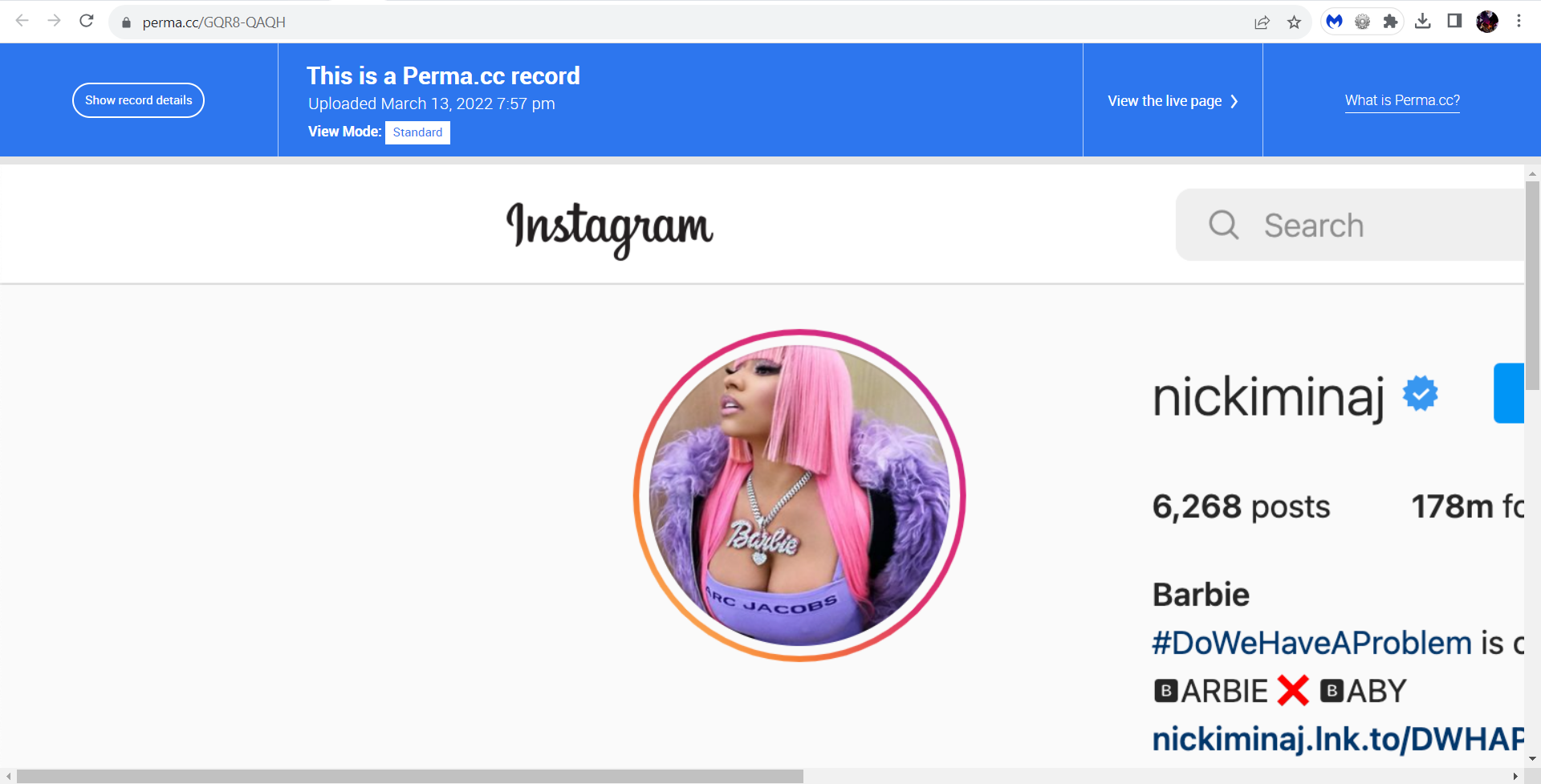}
\caption{Example of an Perma.cc Instagram memento saved as an image}
\label{fig:perma_png}
\end{figure*}

\begin{figure*}[ht]
\centering
\includegraphics[width=.75\linewidth]{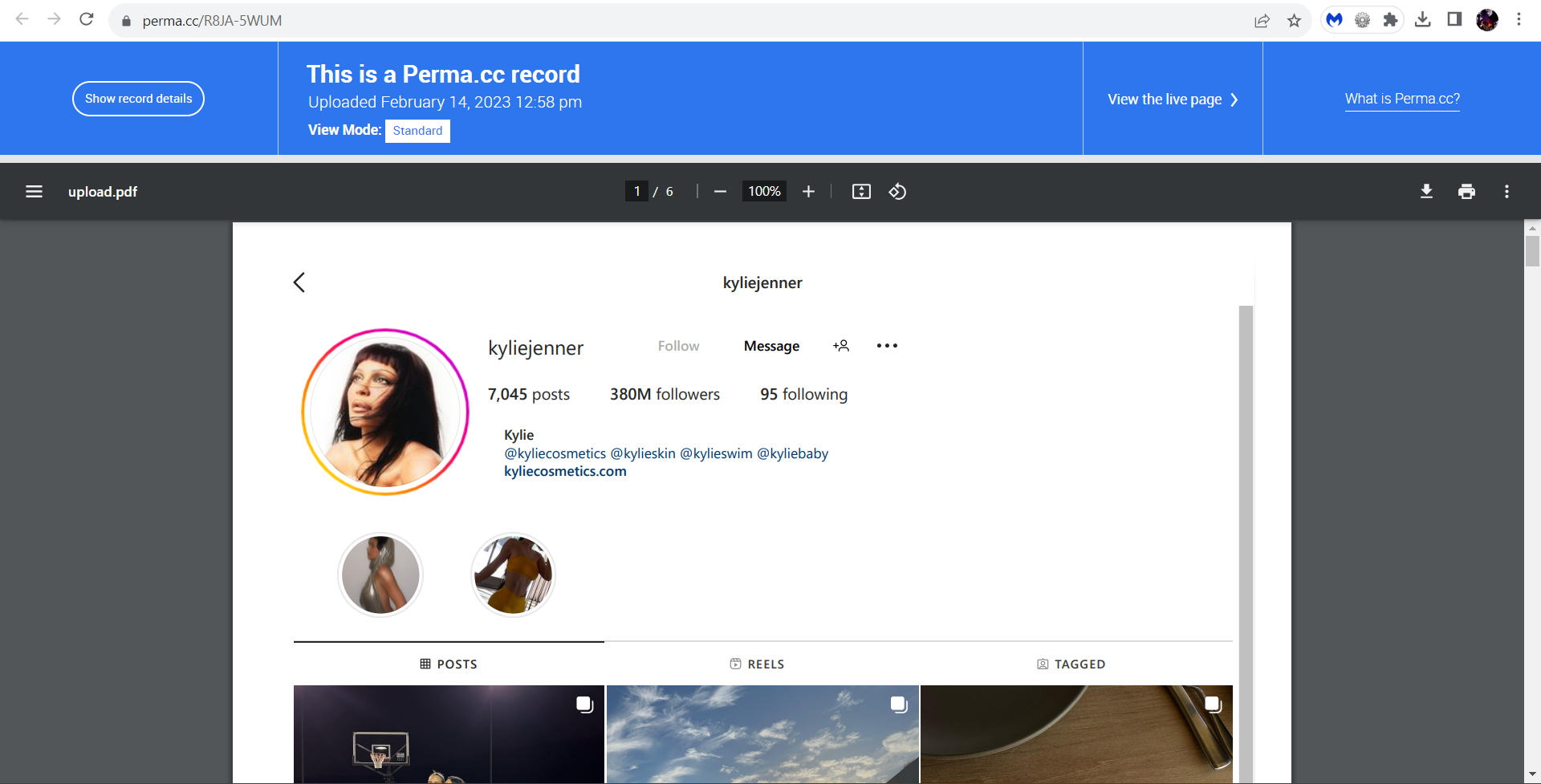}
\caption{Example of an Perma.cc Instagram memento saved as a PDF}
\label{fig:perma_pdf}
\end{figure*}

Perma.cc does not have a public browsing interface nor does it provide a CDX API, making it difficult to collect Instagram mementos from the archive. However, we used MemGator \cite{jcdl-2016:alam:memgator} to retrieve the mementos of the 25 most followed Instagram accounts. From the 32 mementos MemGator returned, only 1 was from 2019. At that rate, one would need at least 2500 Instagram account handles to collect even 100 mementos from 2019. In the end, we decided to only analyze the 32 mementos of the top 25 most followed Instagram accounts. Even from a small amount of mementos, we noticed features of Perma.cc Instagram mementos that were not present in other public archives we studied. For example, if Perma.cc fails to archive an URL, it will give users the option to upload images (Figure \ref{fig:perma_png}) or PDFs of the webpage (Figure \ref{fig:perma_pdf}) instead.\footnote{\url{https://perma.cc/docs}}

\subsection{The Current State of Archiving Instagram}
\begin{figure*}[ht]
\centering
\includegraphics[width=.75\linewidth]{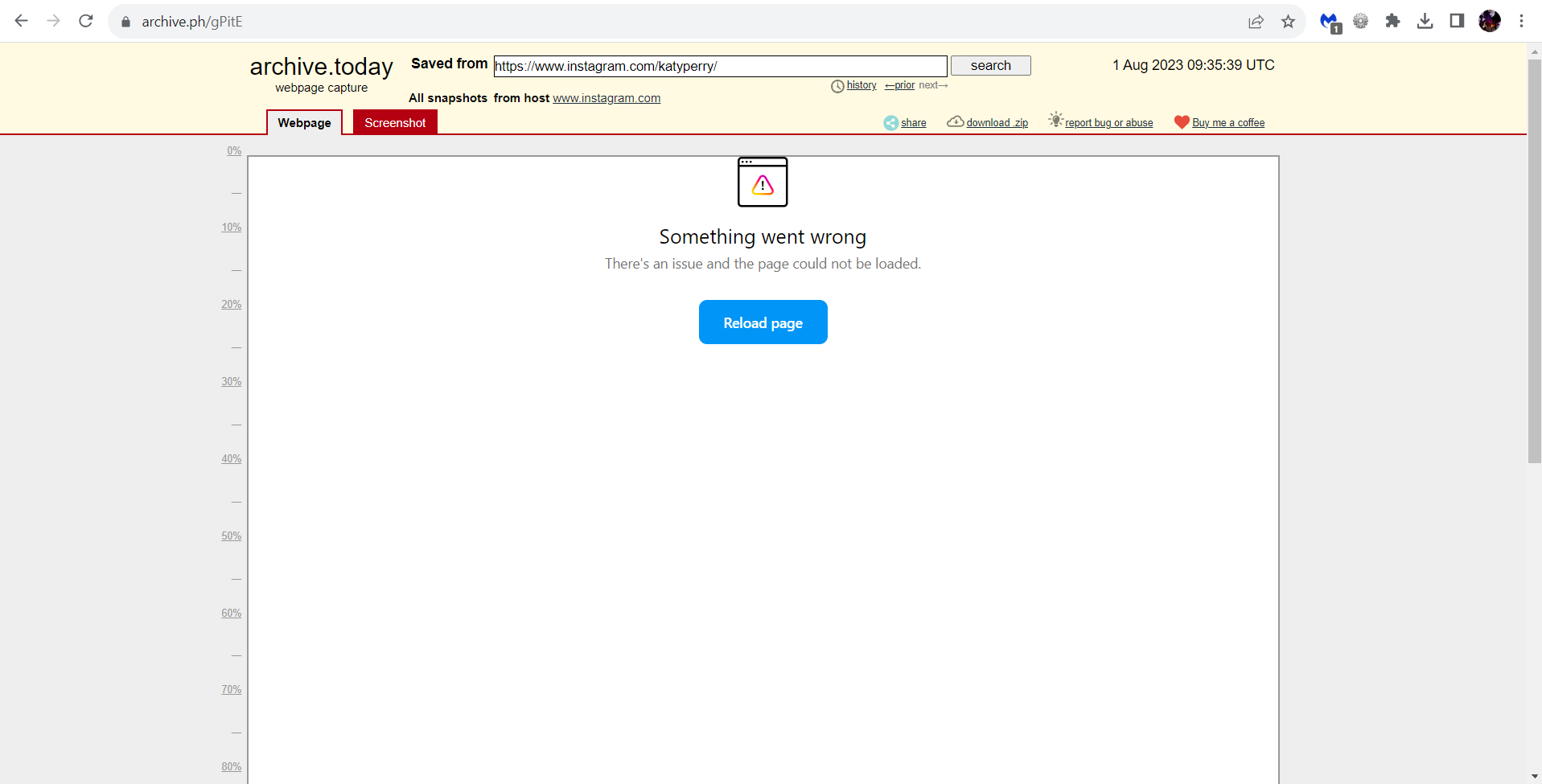}
\caption{Attempt on August 1, 2023 at archiving Katy Perry's Instagram account using Archive.today }
\label{fig:archive_today_archiving_attempt}
\end{figure*}

\begin{figure*}[ht]
\centering
\includegraphics[width=.75\linewidth]{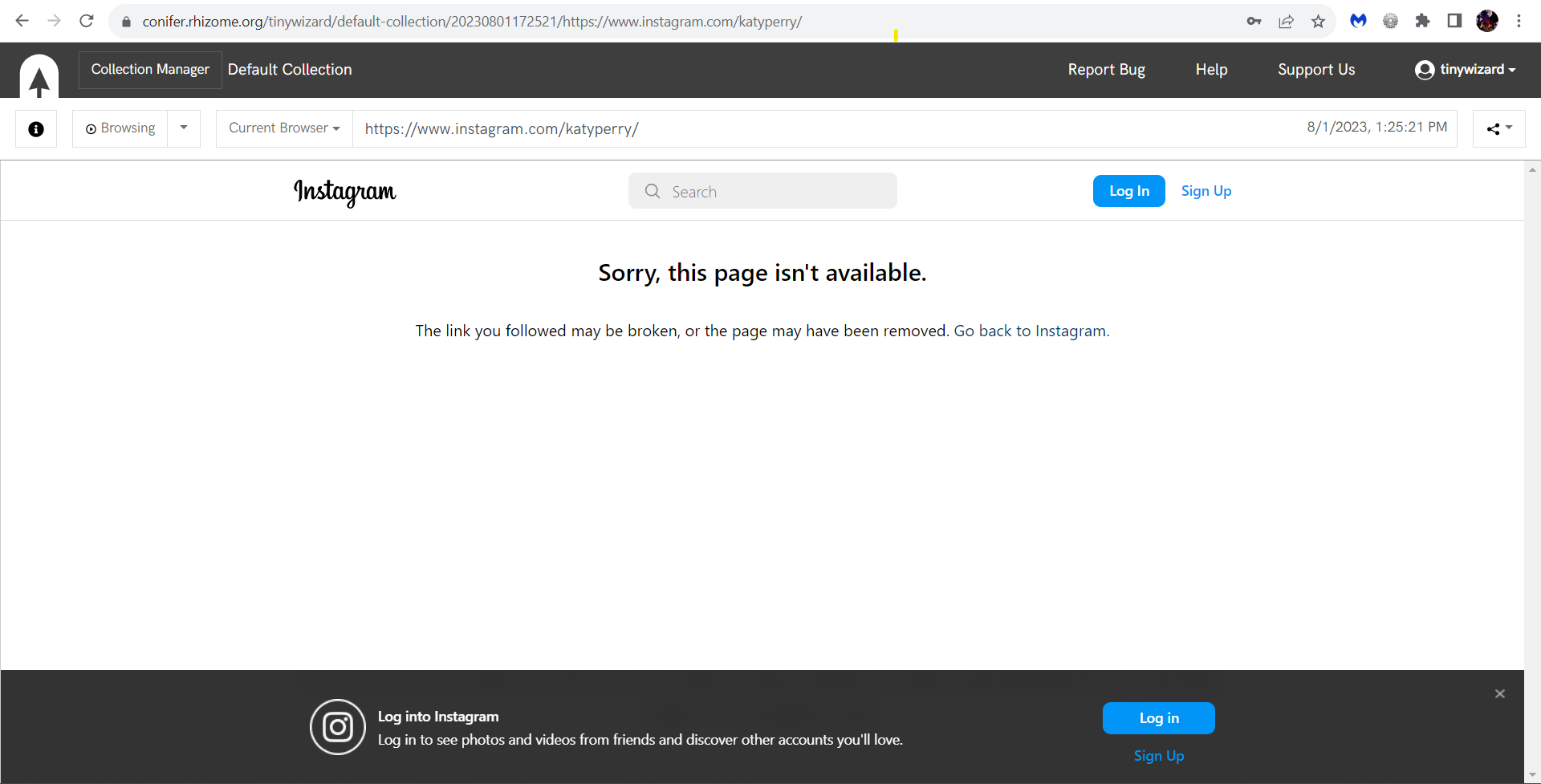}
\caption{Attempt on August 1, 2023 at archiving Katy Perry's Instagram account using Conifer without logging in}
\label{fig:conifer_archiving_attempt_no_login}
\end{figure*}

\begin{figure*}[ht]
\centering
\includegraphics[width=.75\linewidth]{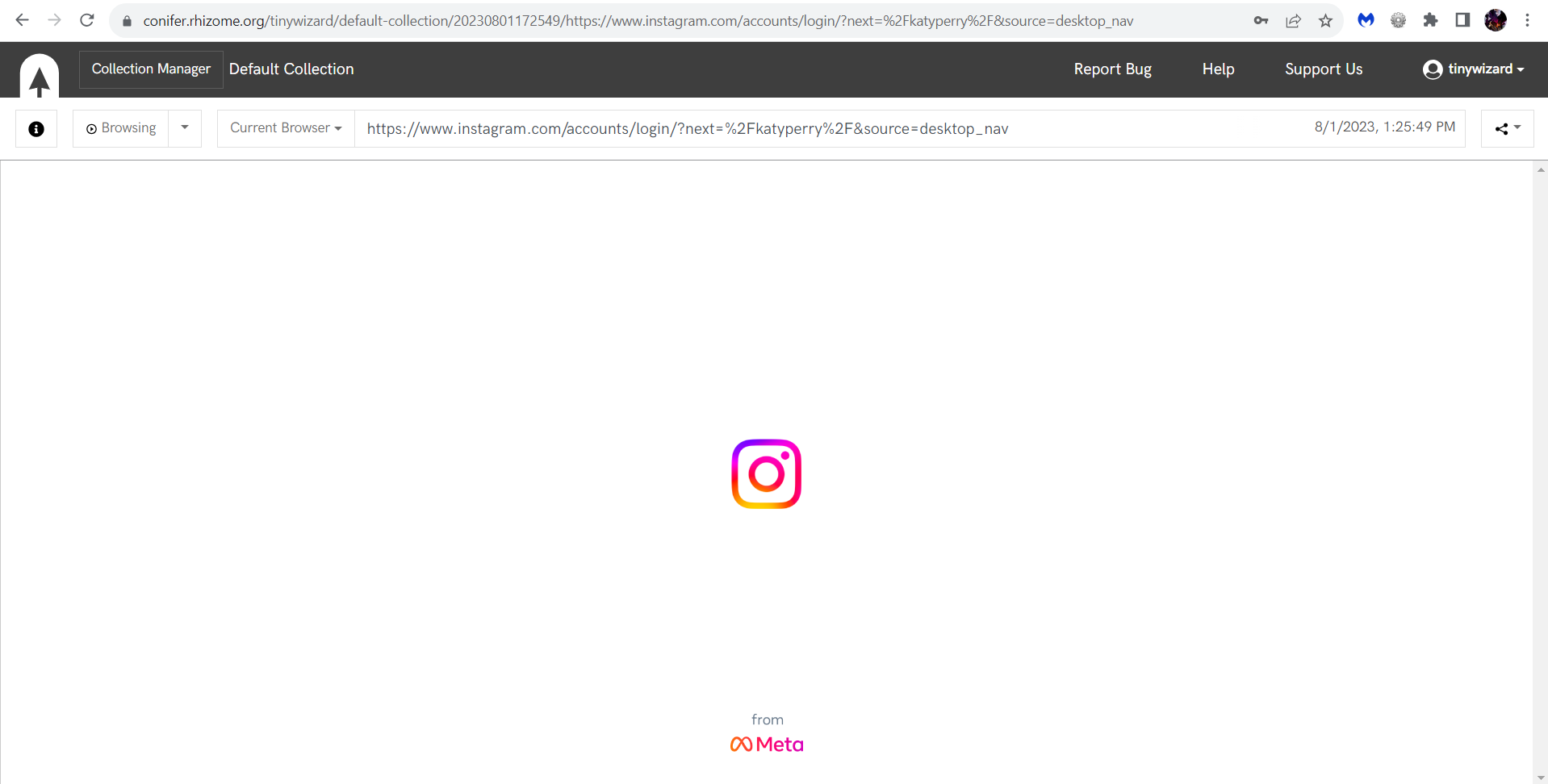}
\caption{Attempt on August 1, 2023 at archiving Katy Perry's Instagram account using Conifer with attempt to log in}
\label{fig:conifer_archiving_attempt_login}
\end{figure*}

\begin{figure*}[ht]
\centering
\includegraphics[width=.75\linewidth]{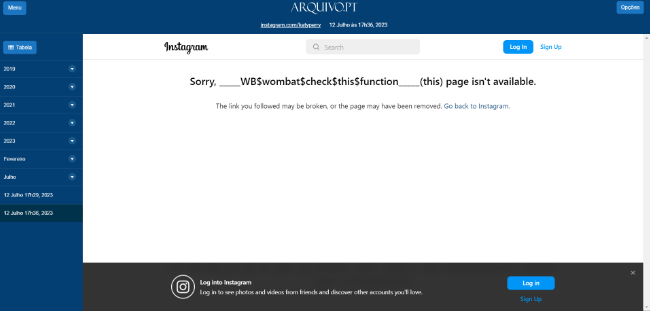}
\caption{First attempt on July 12, 2023 at archiving Katy Perry's Instagram account using Arquivo.pt}
\label{fig:arquivo_archiving_first_attempt}
\end{figure*}

We tried archiving Katy Perry's Instagram account using Archive.today, Arquivo.pt, and Conifer to see if there was a better alternative to the Wayback Machine. We attempted to archive Katy Perry's Instagram account approximately five times. All the attempts resulted in a ``page can't load" message or something similar, as seen in Figure \ref{fig:archive_today_archiving_attempt}. Similarly, none of the five attempts at archiving Katy Perry's Instagram account using Conifer were successful. We also tried to login to our personal Instagram account in Conifer, but that was not successful either (Figures \ref{fig:conifer_archiving_attempt_no_login} and \ref{fig:conifer_archiving_attempt_login}). Lastly, we attempted to archive Katy Perry's Instagram account using Arquivo.pt. In contrast to the Wayback Machine, Arquivo.pt's SavePageNow feature implements a Webrecorder.\footnote{\url{https://webrecorder.net/}} With that in mind, we tried archiving Katy Perry's account with and without logging in. Our first attempt at archiving on July 12, 2023, depicted in Figure \ref{fig:arquivo_archiving_first_attempt} was unsuccessful with and without trying to log in. However, on our second attempt on August 1, 2023, we were able to login into our Instagram account and browse Instagram like a logged in user would be able to. For example, we were able to see the accounts we followed and view the content on Katy Perry's account page. The last archiving attempt has not been integrated into the public collection yet, so we are not able to draw any conclusion on the quality of the memento as of August 10, 2023. 
In summary, none of our attempts at archiving were successful.

\section{Scraping Metadata from Instagram Account Page Mementos}
\begin{figure*}[ht]
\centering
\includegraphics[width=.9\linewidth]{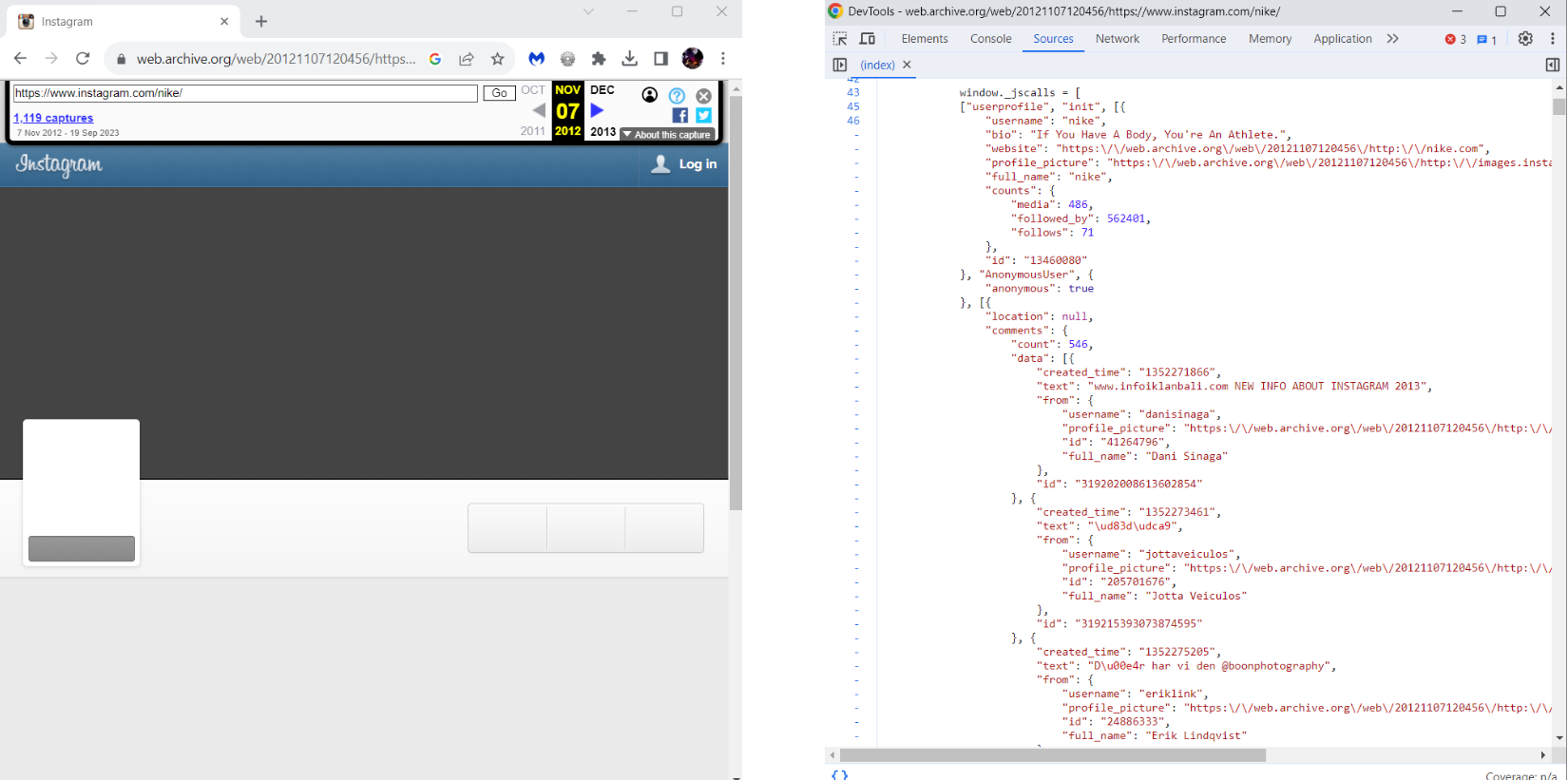}
\caption{The page source of a memento can reveal information that is not visible from the webpage}
\label{fig:blank_memento_with_page_source}
\end{figure*}

 We developed a Python script, instagram\_memento\_scrape.py, using Beautiful Soup\footnote{\url{https://www.crummy.com/software/BeautifulSoup/bs4/doc/}} and the requests\footnote{\url{https://requests.readthedocs.io/en/latest/}} library to scrape data from Instagram account page mementos in the Wayback Machine. As of August 2023, the script can parse Instagram mementos between November 7, 2012 and June 8, 2018. This script is especially useful for retrieving information that is not visible from the replayed webpage of the URI-M. For example, as demonstrated by Figure \ref{fig:blank_memento_with_page_source}, a memento may appear to be completely blank, but its page source can still contain valuable information, such as the follower count, post captions, and image sources.
 
\begin{figure*}[ht]
\centering
\includegraphics[width=.9\linewidth]{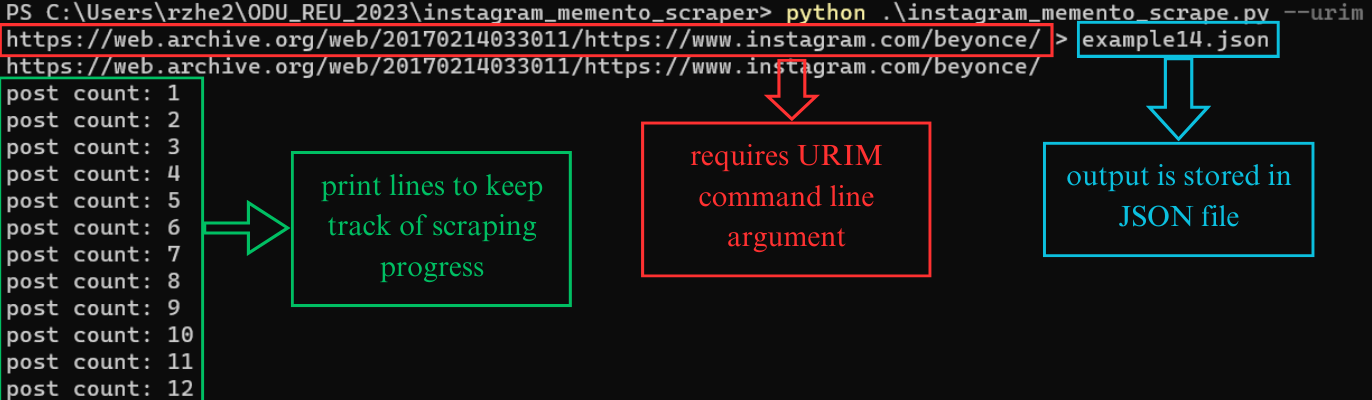}
\caption{Example of using instagram\_memento\_scrape.py}
\label{fig:script_usage}
\end{figure*} 

\begin{figure}
\begin{lstlisting}
{
    "profileUser": {
        "username": "beyonce",
        "bio": "#LEMONADE",
        "website": "https://web.archive.org/web/20170214033011/http://www.beyonce.com/",
        "profile_picture": {
            "uri": "https://web.archive.org/web/20170214033011/https://scontent-sea1-1.cdninstagram.com/t51.2885-19/s150x150/12918039_230227960666719_282379501_a.jpg",
            "status_code": 200
        },
        "full_name": "Beyonc\u00e9",
        "count": {
            "media": 1403,
            "followed_by": 94709950,
            "follows": 0
        },
        "id": "247944034",
        "isVerified": true
    },
    "userMedia": [
        {
            "comments_disabled": false,
            "comments": {
                "count": 504384
            },
            "caption": {
                "text": "We would like to share our love and happiness. We have been blessed two times over. We are incredibly grateful  that our family will be growing by two, and we thank you for your well wishes. - The Carters"
            },
            "short_code": "BP-rXUGBPJa",
            "likes": {
                "count": 10400019
            },
            "created_time": 1485974340,
            "images": {
                "display": {
                    "url": "https://web.archive.org/web/20170214033011/https://scontent-sea1-1.cdninstagram.com/t51.2885-15/e35/16465013_1625467001093055_3757710872030478336_n.jpg?ig_cache_key=MTQ0MDc3OTY0ODkyODkwMzc3MA%3D%3D.2",
                    "status_code": 200
                },
\end{lstlisting}
\caption{A snippet of an example output from the scraper. The full output can be found at \protect \url{https://github.com/oduwsdl/Zheng-REU-2023/blob/main/instagram_memento_scraper/example_outputs/example14.json}.}
\label{fig:example_output}
\end{figure}

The format of the page source of Instagram mementos can often change due to Instagram UI changes, for instance. Because of these frequent changes, we tested our script with at least one Instagram URI-M from every month between November 7, 2012 to June 8, 2018. We created a list of mementos to test by writing a script to select a memento from each unique date from our data set of mementos from the top 25 most followed Instagram accounts. From there, we randomly selected a memento from each month to test. We also tested mementos from dates right before and after a format change in the page source.

\begin{figure*}[ht]
\centering
\includegraphics[width=.9\linewidth]{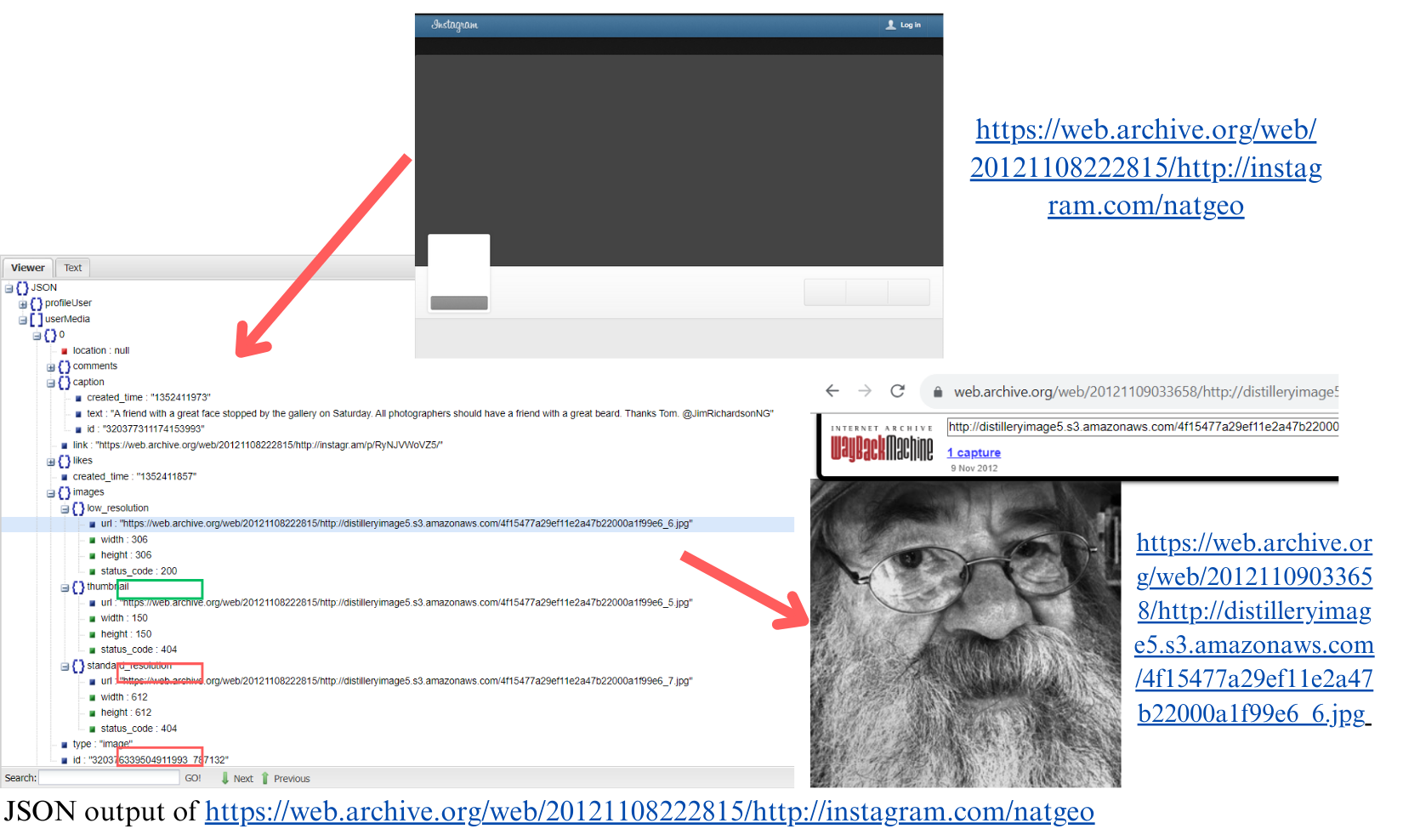}
\caption{You may be able to view an image resource even if the image doesn’t replay on the memento}
\label{fig:why_multiple_image_src}
\end{figure*}

The Python script takes the URI-M to scrape as a command line argument (Figure \ref{fig:script_usage}) and outputs a JSON object (Figure \ref{fig:example_output}). There are two main parts to the JSON output,  profileUser and userMedia. The profileUser section contains basic information about the Instagram user, such as the follower count, username, and biography. The userMedia section is structured by post and contains information about the Instagram user's posts such as the post caption, comment count, and like count. It is important to note that the userMedia section does not contain information on every post the Instagram user has posted, every comment, nor every like but rather just the information present in the page source of the memento. In addition, for any image resource we issue a GET request on the image URL to retrieve the HTTP status code. A status code of 200 indicates that the image resource is archived in the Wayback Machine. Many of the Instagram mementos contain multiple image resources for one post. This means that you may still be able to view an image resource even if the image does not replay on the memento (Figure \ref{fig:why_multiple_image_src}). For a detailed description of the available information for a given URI-M, see Table of Information Extractable from Mementos.pdf.\footnote{\url{https://github.com/oduwsdl/Zheng-REU-2023/blob/main/instagram_memento_scraper}}

Our script's main function is to mine data from the mementos of Instagram accounts. However, in combination with other Python scripts or tools, our scraper could also be used to:
\begin{itemize}
    \firstitem{Perform trend analysis (e.g., follower growth, media count growth)}
    \item Detect hashtags or mentions in post captions or user biography (e.g., to potentially identify backup accounts for disinformation actors or to trace hashtag networks)
    \item Collect all the shortcodes in the mementos of a given Instagram user
    \item Determine if a post, website, or image has been archived
    \item Determine if a post, website, or image is still on the live web, to identify if the post had been deleted
\end{itemize}

\section{Future Work}
Moving forward, we will continue to work on our Instagram memento scraper. First and foremost, we want to expand our Python script to scrape mementos after June 8, 2018. We are also planning to  expand the functionality and usability of our script. Some of the improvements include making the request for the status code of image resources optional, fixing encoding issues with emojis, converting UNIX timestamp to datetime format, and providing more uniformity in the JSON output. Aside from working on \texttt{instagram\_memento\_scrape.py}, we intend to write another Python script to parse the JSON output from our scraper. With these tools, we can do an analysis into how many Instagram account page mementos display as blank but contain content that could be scraped.

\section{Conclusion}
 From this study, we determined that Instagram mementos in the Wayback Machine began redirecting to the Instagram login page in August 2019. This has resulted in a low percentage of replayable mementos in recent years. Since 2020, less than 10\% of the mementos of the top 25 most followed Instagram accounts are replayable. Furthermore, Instagram mementos on other public archives are also not well archived. Without a proper way to archive Instagram pages, we stand to lose valuable data and history.
 However, that is not to say that Instagram mementos are not important or valuable. Instagram mementos are especially helpful in determining changes to an Instagram webpage or viewing content from banned Instagram account. Although there are many issues present in archiving Instagram, there are many replayable Instagram mementos we can extract data from for disinformation research. In fact, from 2012 to 2018 more than 80\% of the mementos of the top 25 most followed Instagram accounts, Disinformation Dozen, and Health Authorities are replayable. We wrote a Python script to scrape information such as follower count and user bio from Instagram mementos in the Wayback Machine. Our Python script performs the crucial first step to studying the content of Instagram mementos which is gathering the content. 
 
\section*{Acknowledgements}
This work was supported by NSF CISE REU Site Award \#2149607.

\renewcommand{\refname}{References Cited}
\bibliographystyle{ieeetr}
\bibliography{refs}
 
\end{document}